*Basov M. V.*
Dukhov Automatics Research Institute

# PRESSURE SENSOR CHIP WITH NEW ELECTRICAL CIRCUIT FOR 10 KPA RANGE

*Abstract:* Characteristics of high sensitivity MEMS pressure sensor chip for 10 kPa utilizing a novel electrical circuit are presented. The electrical circuit uses piezosensitive differential amplifier with negative feedback loop (PDA-NFL) based on two bipolar-junction transistors (BJT). The BJT has a vertical structure of n-p-n type (V-NPN) formed on a non-deformable chip area. The circuit contains eight piezoresistors located on a profiled membrane in the areas of maximum mechanical stresses. The experimental results prove that pressure sensor chip PDA-NFL with 4.0×4.0 mm$^2$ chip area has sensitivity S = 10.1 ± 2.3 mV/V/kPa with nonlinearity of 2$K_{NL}$ = 0.26 ± 0.12 %/FS (pressure is applied from the back side of pressure sensor chip).

## 1. INTRODUCTION

Large percentage of measurements in INDUSTRIAL applications (automotive, process control, aviation, space, medical, power plants) and in R&D (fluid mechanics, robots, biophysics, acoustics, geophysics and others), are linked to measurements of pressure, flow, level, consumption of fluid and made with help of pressure sensors. The pressure sensor market (after first-level packaging) is currently growing with CAGR of 3.8% and will reach $2.0B in 2023 (vs $1.6B in 2016) [1]. Majority of pressure sensors utilize either piezoresistive or capacitive principle of operation. Growing demand for pressure sensors stimulates development of novel types of MEMS pressure sensor chips with improved performance, including higher sensitivity, better accuracy, and smaller die size. The chip area is "main price" which determines sensor output characteristics and its cost. Often piezoresistive pressure sensors assume a similar kind of circuit - this is the Wheatstone bridge. The most influential geometric size of membrane is thickness of thinned membrane part, which will be reduced to a certain limit [2]. Many developments consider the possibility of MS increase in piezoresistor (PR) areas by combinations of mechanical concentrator in the form of rigid islands (RIs) or complex shaped mechanical structures [3–10]. Such mechanical structures are created by using deep reactive ion etching (DRIE) of both back and top side of pressure sensor chip. The use of DRIE is rather expensive process in mass production. Another option for improving pressure sensitivity is related to introducing active elements as transistors to electrical circuit formed on pressure sensor chip. The proof of this method will be presented below for differential pressure range of 10 kPa.

## 2. STRUCTURE AND MODELING

This research shows us, that sensitivity can be increased due to the use of new electrical circuit utilizing bipolar junction transistor (BJT) at MEMS pressure sensor chip operating for pressure range of -10...+10 kPa. The pressure sensor chip utilizes electrical circuit with BJT-based piezosensitive differential amplifier with negative feedback loop (PDA-NFL) [11-17]. Advantage of the selected electrical circuit relative to classic Wheatstone bridge circuit is in the possibility of combining of a larger number of PRs (8 instead of 4) because BJTs are three-pole devices (Fig. 1). The certain combination of nominal values of PRs and their dependence on mechanical stress allows for larger changes of BJT collector potential. The difference between these potentials is used as the circuit output signal. The NFL in electrical circuit allows for reduction of the temperature dependence of the BJT base potential so temperature dependence of the output signal can be reduced. The circuit works at the same supply voltage $U_{sup} = 5.0$ V as it usually uses for Wheatstone bridge.

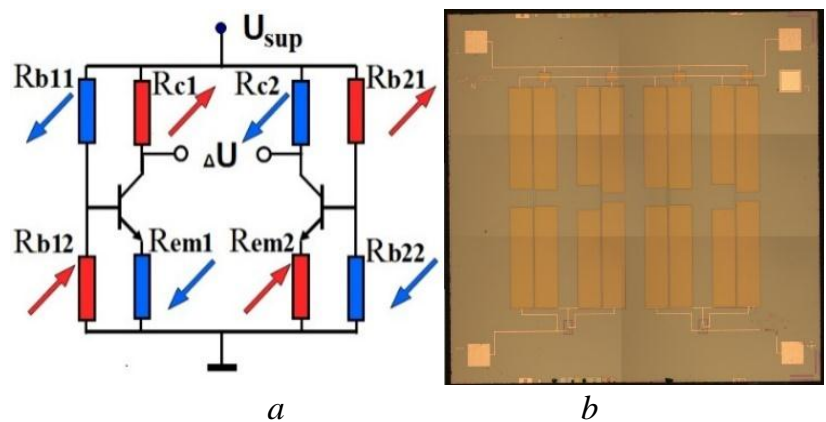

*Fig. 1*. Pressure sensor chip PDA-NFL:
*a* – electrical circuit with indicators of change direction for PRs when pressure is applied from the back side of chip; *b* – top view of chip

At the design phase, electrical circuits were studied using both analytical methods and modelling in NI Multisim software (Fig.2). Dependence of die output signal on pressure ΔU(ΔP) was derived based on known values of piezoresistive coefficients, data on mechanical stress distribution obtained with ANSYS (Fig. 3) and results of modelling of I-V characteristic obtained in Synopsys TCAD software (Fig. 4).

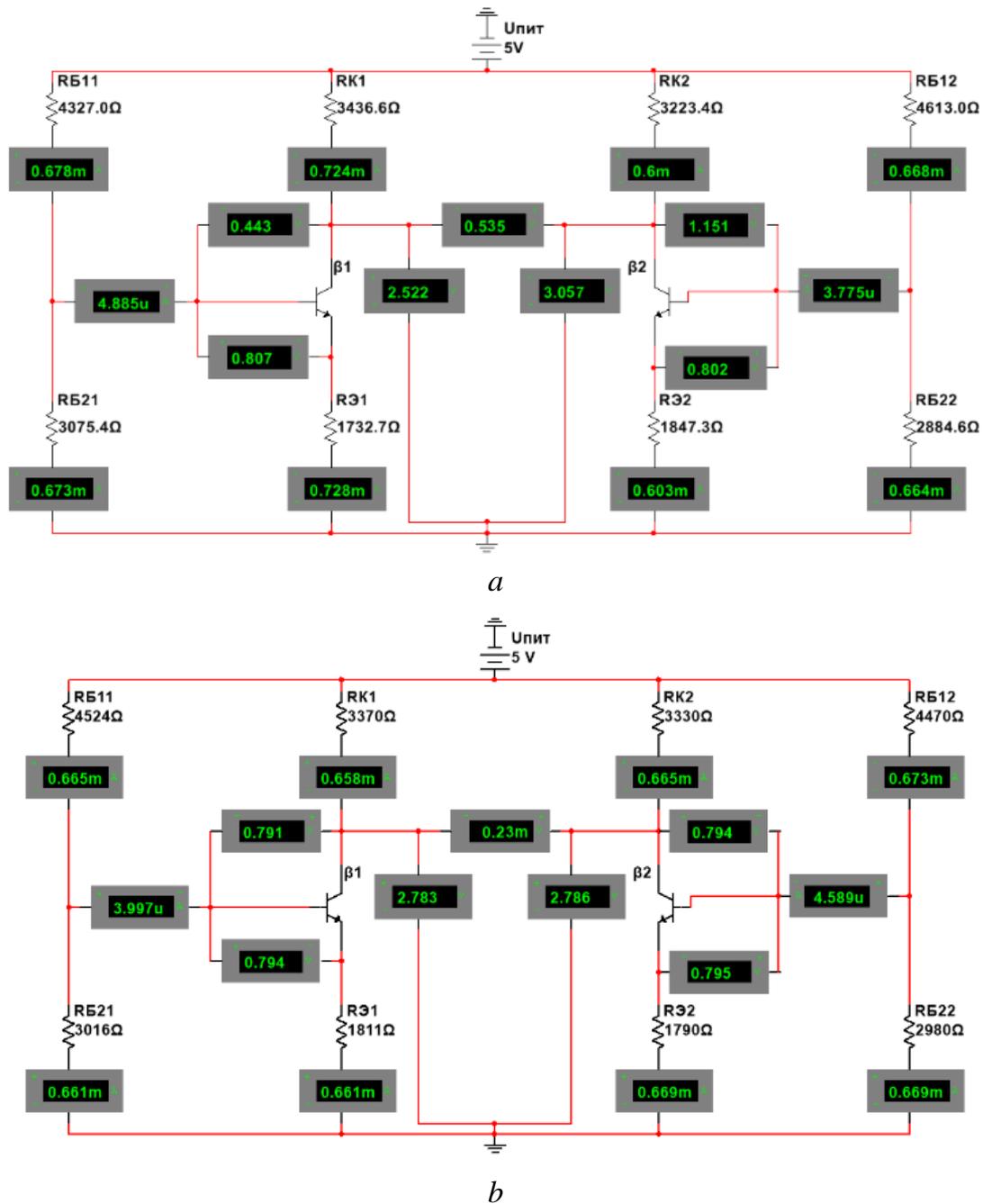

*Fig. 2.* Values of electrical circuit components for PDA-NFL chip loaded by pressure and temperature from NI Multisim software

It is necessary to define some input parameters for analysis of BJT, because the operation point of BJT for developed circuit must found. Geometrical parameters of membrane are shown in Fig. 3 and Table 1. Target values of piezoresistors and operating points of BJTs are listed in Table 2.

NI Multisim software was used for a detailed analysis of circuit parameters. The circuit was modelled by combining discrete elements with predetermined parameters. Gummel-Poon model was used for BJT description. Reverse current of collector junction was calculated based on analysis of a stand-alone transistor with a requirement of getting base current of $I_b = 0.3$ µA at a given emitter-base voltage $U_b$-$U_{em}$.

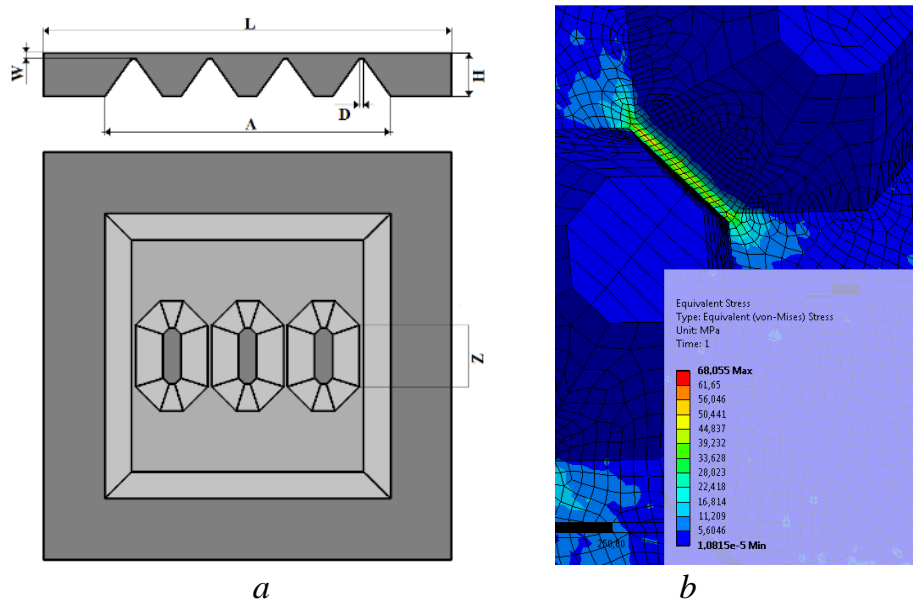

*Fig. 3.* Back view of pressure sensor chip PDA-NFL and schematic image for cut of chip mechanical part with geometrical indicators and distribution of the mechanical stress (von Mises by ANSYS) between rigid islands (RI)

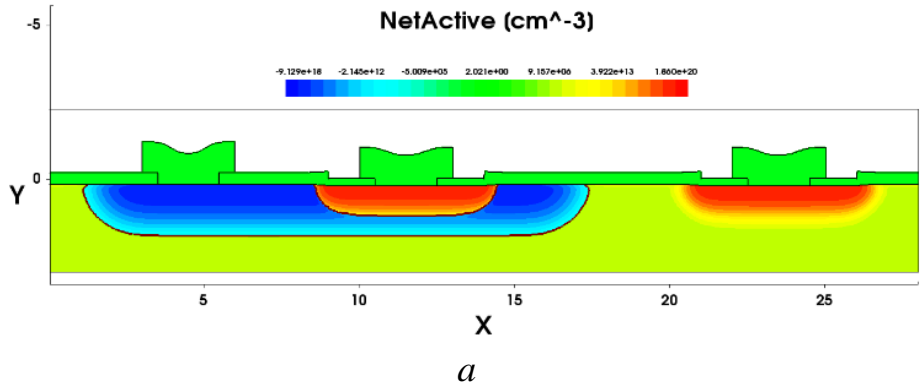

*a*

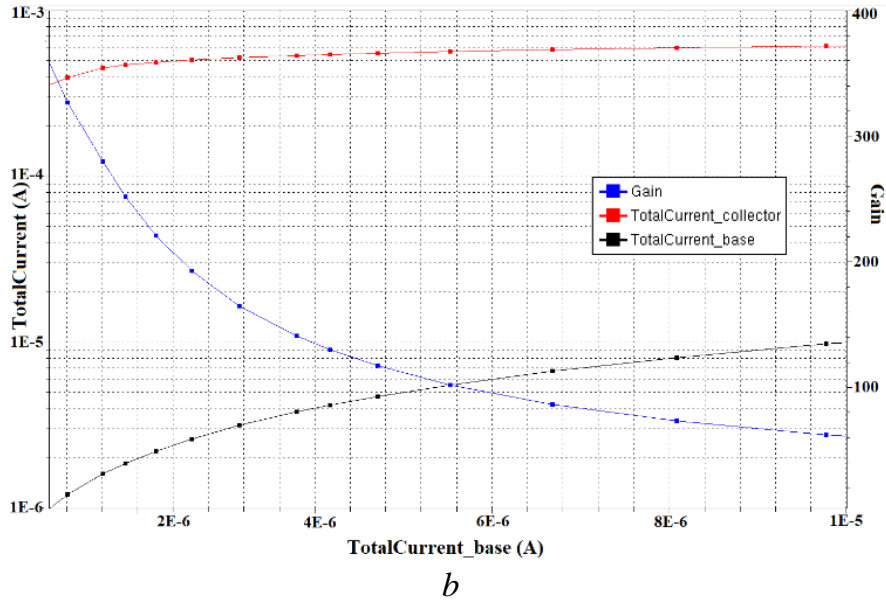

*b*

*Fig. 4.* Structure of BJT V-NPN transistor (a) and dependence of gain β on base current (b) in Synopsys TCAD software

Modelling results for electrical circuit when sensor chip is loaded by $\Delta P = 10$ kPa (Fig. 2a) pressure and temperature $\Delta T = 10\ ^0C$ (Fig. 2b) is presented. To illustrate temperature dependence of circuit parameters, left side of the circuit is shown with parameters affected by temperature increase by $10\ ^0C$ ($T_1 = 37\ ^0C$) while the right side shows circuit parameters at temperature $T_{room} = 27\ ^0C$.

*Table 1*

Geometrical Parameters of Pressure Sensor Chip PDA-NFL

| **Geometrical Parameter** | **Size, μm** |
|---|---|
| Chip side L (square) | 4000 |
| Thickness of tinned membrane part W | 12 |
| Thickness of thickened membrane part H | 410 |
| Membrane side A | 2280 |
| Groove width between RIs or RI and chip frame D | 28 |
| Length of RI edge Z | 680 |

Parameters of PDA-NFL Circuit

| Parameters | | Value |
|---|---|---|
| **Bipolar junction transistor** | Base current $I_b$, µA | 4.6 |
| | Gain β | 145 |
| | Emitter-base voltage drop $U_{em-b}$, V | 0.80 |
| | Collector-base voltage drop $U_{c-b}$, V | 0.80 |
| | Collector potential $U_c$, V | 2.79 |
| **Resistors** | $R_{b11, b21}$, KOhm | 4.5 |
| | $R_{b12, b22}$, KOhm | 3.0 |
| | $R_{c1, c2}$, KOhm | 3.3 |
| | $R_{em1, em2}$, KOhm | 1.8 |

## 3. FABRICATION PROCESS

Pressure sensor chips were manufactured using (100) p-type Si wafers (diameter 3 inches) with n-type epitaxial layer. It is necessary for separating the BJT regions on chip area. Isolation diffusion areas ($p^+$) extending through full thickness of epi-layer are used to electrically isolate BJTs and PRs. P-type substrate is connected to ground. PRs contain high-doped $p^+$-type regions ($N_{sp+} = 7.4 \cdot 10^{19}$ cm$^{-3}$, $x_{jp+} = 3.6$ µm, $R_{sp+} = 17$ Ohm/cm$^2$) connecting the bond pads located on the frame with low-doped p-type regions described early. The closest metallization region of aluminum with thickness of 0.8 µm is located at minimum distance of 80 µm from the thinned part boundary of membrane. It is necessary to reduce negative properties, which were associated with a significant difference between the linear coefficients of temperature expansion for silicon and aluminum. Although this type of substrates allows for electrochemical etching with etch stop and p-n junction, that option was not used for membrane fabrication. Membranes formed using timed wet etching had significant thickness variation. Consequently, output parameters also had large variation. In further studies, preferably, mechanical structure should have membranes formed using micromachining process with etch stop – either DRIE or wet stop etching [18-22]. Technological route for processing pressure sensor wafers: 1. Oxidation; 2. The sequence of cleaning, photolithography and doping steps, including: a. boron for isolation areas and creating contact to the substrate, b. boron for high-doped PR regions, c. boron for low-doped PR regions and BJT base areas, d. phosphorous for BJT emitter and collector areas; 3. $Si_3N_4$ deposition as protection layer for membrane etching; 4. Cavity photolithography on the back side of the wafer; 5. Wet anisotropic etching of membrane in 30% KOH aqueous solution at T = 85 °C; 6. Wet isotropic etching of membrane in a mixture of HF : $HNO_3$ : $CH_3COOH$ (2:9:4); 7. Removal of $Si_3N_4$ layer; 8. Photolithography and etching to open contacts; 9. Sputtering of Al-Si (1.5%); 10. Photolithography and metal etching to define metal lines and bond pads; 11. Dicing. Pressure sensor chip PDA-NFL was bonded to a silicon support and placed on Kovar case (Fig. 5) for testing of its characteristics.

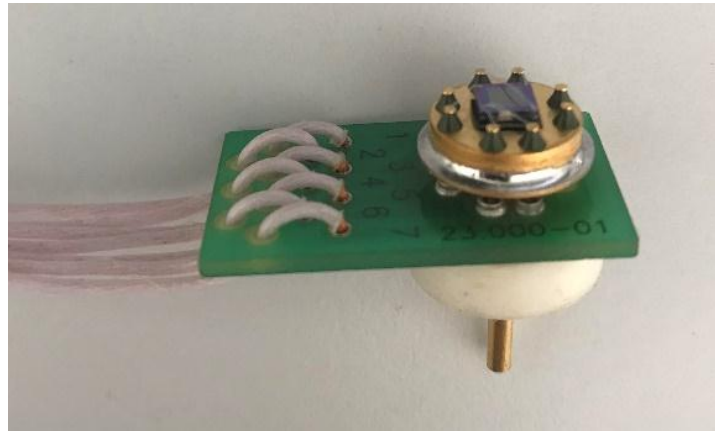

*Fig. 5.* Assembly of pressure sensor PDA-NFL

The silicon support consists of an inter-mediate element and a pedestal, which are connected by low-temperature glass. The sensor assembly allows for measuring differential pressure. All samples were exposed to temperature and pressure cycling to remove residual assembly stress before functional testing [23-28].

## 4. COMPARATIVE ANALYSIS

18 samples of pressure sensors with PDA-NFL chip were compared with similar products utilizing Wheatstone bridge circuit having the same pressure range of 10 kPa. As an analogue, a pressure sensor chip with an area of 6.15x6.15 $mm^2$ is used, which is 2.36 times more than in the proposed development. This analogue is produced on the same technological line as the PDA-NFL chip. Sensitivity and related parameters reported from measurements where pressure was applied from the both sides. The values of the zero output signal, sensitivity and nonlinearity are presented in Table 3.

*Table 3*

Parameters of Zero Output Signal, Sensitivity and Nonlinearity

| The circuit of Chip | | | | PDA-NFL | Wheatstone bridge |
|---|---|---|---|---|---|
| Parameters | | Symbol | Units | | |
| Sensitivity (pressure from the back side) | Change of signal | $\Delta U_{10}$ | mV | 505.5 ± 115.6 | 87.0 ± 14.7 |
| | Relative value | S | mV/V/kPa | 10.11 ± 2.31 | 1.74 ± 0.29 |
| Sensitivity (pressure from the front side) | Change of signal | $\Delta U_{10}$ | mV | 513.9 ± 115.6 | 88.8 ± 15.0 |
| | Relative value | S | mV/V/kPa | 10.28 ± 2.31 | 1.78 ± 0.30 |
| Nonlinearity | pressure from the back side | $2K_{NL}$ | %/10 kPa | 0.27 ± 0.13 | 0.04 ± 0.02 |
| | pressure from the front side | | | 0.25 ± 0.12 | 0.03 ± 0.02 |
| Limit of zero output signal | | $U_0$ | mV/V | < 3 | < 6 |

Temperature characteristics were measured in two temperature ranges: from -55 °C to +25 °C and from +25 °C to + 55 °C for temperature hysteresis; from +5 °C to +25 °C and from +25 °C to + 45 °C for temperature coefficient of zero signal or sensitivity (see Table 4).

*Table 4*

Parameters of Temperature Characteristics

| The circuit of Chip | | | | PDA-NFL | Wheatstone bridge |
|---|---|---|---|---|---|
| Parameters | | Symbol | Units | | |
| Temperature hysteresis of zero signal -55…+25 °C | pressure from the back side | THZ | % | 0.09 ± 0.04 | 0.07 ± 0.07 |
| | pressure from the front side | | | 0.09 ± 0.04 | 0.07 ± 0.07 |
| Temperature hysteresis of zero signal +25…+55 °C | pressure from the back side | | | 0.12 ± 0.05 | 0.04 ± 0.03 |
| | pressure from the front side | | | 0.12 ± 0.05 | 0.04 ± 0.03 |
| Temperature hysteresis of sensitivity -55…+25 °C | pressure from the back side | THS | | 0.11 ± 0.08 | 0.28 ± 0.05 |
| | pressure from the front side | | | 0.11 ± 0.07 | 0.29 ± 0.04 |
| Temperature hysteresis of sensitivity +25…+55 °C | pressure from the back side | | | 0.10 ± 0.08 | 0.28 ± 0.06 |
| | pressure from the front side | | | 0.10 ± 0.09 | 0.27 ± 0.04 |
| Temperature coefficient of zero signal +5…+25 °C | pressure from the back side | TCZ | %/10ºC | 0.16 ± 0.05 | 0.14 ± 0.13 |
| | pressure from the front side | | | 0.15 ± 0.05 | 0.14 ± 0.13 |
| Temperature coefficient of zero signal +25…+45 °C | pressure from the back side | | | 0.12 ± 0.06 | 0.10 ± 0.09 |
| | pressure from the front side | | | 0.12 ± 0.06 | 0.10 ± 0.09 |
| Temperature coefficient of sensitivity +5…+25 °C | pressure from the back side | TCS | | 2.25 ± 0.15 | 2.36 ± 0.05 |
| | pressure from the front side | | | 2.29 ± 0.27 | 2.35 ± 0.03 |
| Temperature coefficient of sensitivity +25…+45 °C | pressure from the back side | | | 2.22 ± 0.51 | 2.04 ± 0.03 |
| | pressure from the front side | | | 2.16 ± 0.49 | 2.03 ± 0.03 |

Overload pressure (proof pressure) testing was done from both sides at 60 kPa. Stability of output signal was tested for 9 hours (see Table 5).

*Table 5*

Parameters of Stability and Influence of Overload Pressure

| The circuit of Chip | | Symbol | Units | PDA-NFL | Wheatstone bridge |
|---|---|---|---|---|---|
| Parameters | | | | | |
| Stability of zero output signal | pressure from the back side | $dU_{st\,z}$ | %  | 0.02 ± 0.02 | 0.02 ± 0.01 |
| Stability of sensitivity | pressure from the back side | $dU_{st\,s}$ | | 0.03 ± 0.03 | 0.03 ± 0.02 |
| Influence of overload pressure | pressure from the back side | $dU_{over}$ | | 0.01 ± 0.01 | 0.04 ± 0.04 |
| | pressure from the front side | | | 0.01 ± 0.01 | 0.02 ± 0.02 |

The output signal noise was estimated as RMS value of output voltage fluctuation for 5 minutes (see Fig. 6, measurements each 1 second).

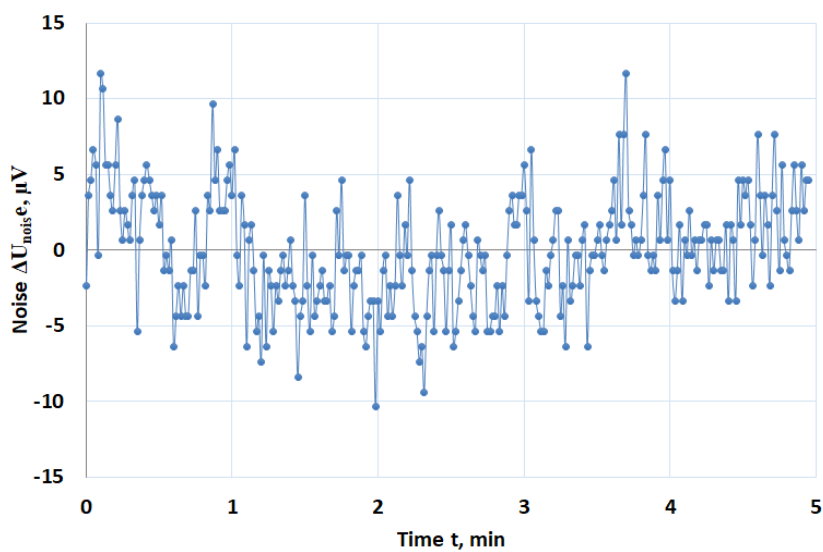

*Fig. 6.* Noise of output signal for pressure sensor PDA-NFL

## 5. REFERENCES

The development of new type of pressure sensor chip using the original electrical circuit PDA-NFL with bipolar transistor V-NPN is presented. Based on the constructed theoretical model the pressure sensor chip was technologically implemented to measure the differential pressure range of 10 kPa. The results of comparative analysis for the development and analogue with Wheatstone bridge proved advantages while maintaining the supply voltage for the electrical circuit (5V) and the range of measured pressure (10 kPa):

1. Sensitivity is increased by 5.8 times;

2. Chip area is reduced by 2.4 times;

3. All parameters of temperature characteristics have the same low range of values. 0.3% for hysteresis and 0.3%/10°C for coefficient;

4. Stability for 9 hours is the same for two types of chips up to 0.05%;

5. The influence of overload pressure is not more than 0.05% for two types.

6. The ratio between signal to noise is more than $3.5 \times 10^4$ in two cases.

There are minor disadvantages in terms of nonlinearity error and zero output signal, but it is significantly critical for the current situation among the performance of other pressure sensor chips [29-38].